\documentstyle[twocolumn,prb,aps,epsf,rotate]{revtex}
\large
\newcommand{\epsv}{\mbox{\boldmath{$\epsilon$}}}
\begin{document}
\epsfverbosetrue
\draft
\twocolumn[\hsize\textwidth\columnwidth\hsize\csname @twocolumnfalse\endcsname
\title
{Orbital Ordering in the Manganites :- Resonant X-ray Scattering Predictions
at the Manganese ${\boldmath L}_{\boldmath II}$ and 
${\boldmath L}_{\boldmath III}$ Edges.}
\author{C.W.M. Castleton$^{1}$, M. Altarelli$^{2,3}$.}
\address{1. European Synchrotron Radiation Facility,
B.P. 220, F-38043 Grenoble C\'{e}dex, France.}
\address{2. Sincrotrone Trieste, Area Science Park, 
34012 Basovizza, Trieste, Italy.} 
\address{3. Abdus Salam International Centre for Theoretical Physics, Strada
Costiera 11, 34014 Trieste, Italy.}
\date{\today} 
\maketitle 

\begin{abstract}
It is proposed that the observation of orbital ordering in manganite materials 
should be possible at the $L_{II}$ and $L_{III}$ edges of manganese using X-ray 
resonant scattering. If performed, dipole selection rules would make
the measurements much more direct than the disputed observations at the 
manganese $K$ edge. They would yield specific
information about the type and mechanism of the ordering not available at
the $K$ edge, as well as permitting the effects of orbital ordering and 
Jahn-Teller ordering to be detected and distinguished from one 
another. Predictions are presented based on atomic multiplet calculations, 
indicating distinctive dependence on energy, as well as on polarization and on 
the azimuthal angle around the scattering vector.

\end{abstract}
\pacs{PACS numbers: 75.30.Vn, 61.10.Dp, 71.70.Ej, 71.70.ch}
]

\section{Introduction.}

The manganite materials, such as La$_{1-x}$Sr$_{x}$MnO$_{3}$ and 
La$_{1-x}$Sr$_{1+x}$MnO$_{4}$, have received much attention recently, due to 
the complex interplay of electronic, spin and orbital degrees of freedom which 
they exhibit. This includes observation of 
colossal magneto-resistance and a large variety of phase transitions as a 
function of temperature, magnetic field and doping. Among the most interesting 
of late have been the charge and orbitally ordered states observed in a variety 
of materials such as La$_{0.5}$Sr$_{1.5}$MnO$_{4}$\cite{Sternleib,Murak}, 
LaMnO$_{3}$\cite{Murak2}, La$_{0.5}$Ca$_{0.5}$MnO$_{3}$\cite{Radaelli}, 
(see also\cite{Vogt,Zimmermann}), La$_{0.33}$Ca$_{0.67}$MnO$_{3}$\cite{Chen}, 
and La$_{0.25}$Ca$_{0.75}$MnO$_{3}$\cite{Cheong}. 
As the temperature is lowered all of the materials (except for the undoped
LaMnO$_{3}$) show a charge ordering transition in which separate sublattices 
develop for Mn$^{3+}$ and Mn$^{4+}$ ions. An orbital ordering transition on the 
Mn$^{3+}$ sublattice (all Mn sites in the case of LaMnO$_{3}$) is then believed 
to occur, followed at (generally) lower temperatures
by a magnetic ordering transition. The structure of all of these orbitally 
ordered states is believed to be very similar, and our results will be relevant 
to all. The exception will be LaMnO$_{3}$, for which the period of the orbital
order is too small, (see later). For simplicity we will refer mostly to the 
layered material, La$_{0.5}$Sr$_{1.5}$MnO$_{4}$, returning to the others at the 
end.
 
In the case of La$_{0.5}$Sr$_{1.5}$MnO$_{4}$ the 
charge ordering transition is at about $T_{CO}=220K$, with a unit variation of 
valence observed between the sublattices\cite{Sternleib,Murak}.
This results in a doubling of the unit cell and the 
appearance of forbidden reflections at, for example, 
$(\frac{\rm 1}{\rm 2},\frac{\rm 1}{\rm 2},0)$. At about $T_{N}=160K$, as seen 
by neutron scattering \cite{Sternleib}, a complex
antiferromagnetic ordering occurs, involving both manganese sublattices. (See 
figure 1.) However, the antiferromagnetic transition observed in the magnetic 
susceptibility\cite{Damay} is higher, concurrent with the charge ordering.
It seems likely, therefore, that in-plane antiferromagnetic order develops at 
a temperature $T_{N(ab)}=220K$ and becomes fully three dimensional at 
$T_{N(c)}=160K$. (Rod-like neutron scattering has been reported between $T_{N}$ 
and $T_{CO}$\cite{Sternleib}.)

At the Mn$^{3+}$ sites the Hund's rule coupling 
is strong, and the crystal field has a large cubic ($O_{h}$) component. 
Described at one electron level, the Mn$^{3+}$ $3d^{4}$ configuration 
thus becomes a two fold degenerate $t_{2g\uparrow}^{3}e_{g\uparrow}^{1}$ 
configuration. (See figure 2.) This degeneracy can be lifted, with (in 
principle) an associated a Jahn-Teller (JT) distortion of the oxygen 
octahedron, reducing the symmetry to $D_{4h}$. Hence we shall denote the two 
components of the
$e_{g\uparrow}$ level as $3d_{3z^2-r^2\uparrow}$ and $3d_{x^2-y^2\uparrow}$.
Goodenough\cite{Good} showed that the spin
ordering is actually dependant upon the ordering of this orbital degree of 
freedom. Above $T_{CO}$ all Mn sites have $3d_{3z^2-r^2\uparrow}$ (part) 
filled, oriented along the crystal $c$ axis with a macroscopic tetragonal 
distortion. Below $T_{N(c)}$, however, a distinctive ``herring-bone'' pattern 
is required in order to explain the observed spin structure, as shown in figure 1. 
This orbital pattern again doubles the 
unit cell, having the wavevector $(\frac{\rm 1}{\rm 4},\frac{\rm 1}{\rm 4},0)$.
This is claimed to have 
been observed recently using resonant X-ray scattering at the Mn $K$ 
edge\cite{Murak}. These results indicate that orbital order develops at the 
same temperature as the charge ordering:- 
$T_{OO} = T_{CO} ( = T_{N(ab)} ) = 220K$. The fact that the spins do not order 
out of plane until a lower temperature $T_{N(c)}$ is not in disagreement with 
this, since for La$_{0.5}$Sr$_{1.5}$MnO$_{4}$ Goodenough's 
orbitally mediated spin interactions only produce couplings in the $ab$ plane, 
not up the $c$ axis. This leaves us with at least two possible mechanisms
for the orbital ordering - it could be due to the spin ordering it permits, or 
to the JT distortions, or to a combination of the two. The question of which 
mechanism is the more important is still disputed. 
In other materials (such as\cite{Rodriguez} LaMnO$_{3}$ and\cite{Radaelli}
La$_{0.5}$Ca$_{0.5}$MnO$_{3}$) the ordering of the JT distortions 
around the Mn$^{3+}$ sites has been observed directly, using high resolution 
neutron and X-ray diffraction, and crystallographic refinement.
The level of distortions appears to vary somewhat, from about 7$\%$ to 12$\%$,
suggesting that the JT mechanism may at least be not the sole mechanism of 
importance. Indeed, in La$_{0.5}$Sr$_{1.5}$MnO$_{4}$ only a 1$\%$ oxygen 
breathing mode has so far been observed\cite{Sternleib}, although detailed 
crystallographic refinement is not reported. It could thus be 
suggested\cite{Murak} that here the JT distortions actually remain along the 
$c$ axis even when the orbitals have ordered in the $ab$ plane, and that the 
only mechanism of importance for this material is the Goodenough spin ordering 
mechanism. The complete absence of accompanying JT distortion ordering in the 
$ab$ plane seems very unlikely, however. More detailed crystallographic 
refinement might be able to clarify this, as was the 
case\cite{Radaelli,Radaelli2} for La$_{0.5}$Ca$_{0.5}$MnO$_{3}$.

What is clear is that the interaction and interdependence of the 
spin, orbital and JT ordering is complex, and not yet fully understood.
In order to approach a better understanding it would be very helpful to be able 
to observe the JT and orbital ordering independently of one another. The $K$
edge experiments so far performed fail to do this.
They are indirect, in the sense that they probe primarily the 
$4p$ shell, rather than the $3d$ shell
in which the supposedly ordered orbitals lie. The sensitivity was thought to
have arisen from a mixture of the Coulomb interaction with the ordered $3d$ 
electrons and the JT distortion of the site\cite{Ishihara}. It has since been 
shown\cite{Elfimov,Benfatto} 
that the experiment is about 100 times more sensitive to the accompanying
JT ordering than to the orbital ordering. Although
an interference term between the two\cite{Benfatto} does leave the
possibility of distinguishing them by looking at the energy dependence of the 
peak, it seems rather doubtful that a direct observation of orbital ordering, 
as distinct from JT ordering, is possible at the Mn $K$ edge. Since the orbital 
believed to order is the Mn$^{3+}$ $3d_{3z^2-r^2\uparrow}$, it seems logical to 
try resonant scattering at the Mn $L_{II}$ and $L_{III}$ edges, probing the 
$3d$ shell itself. Unfortunately these edges lie in the soft X-ray region, 
so, although the Bragg angle for the 
$(\frac{\rm 1}{\rm 4},\frac{\rm 1}{\rm 4},0)$ reflection is real 
($62.9^{\circ}$ at the $L_{III}$ edge), the penetration depth will be very 
short. This will make the experiment surface sensitive and rather difficult, 
but not necessarily impossible. It is certainly the correct way to proceed if 
one wishes to directly probe the orbital order in these materials.

In the next section we will discuss the origin of the scattering and its
azimuthal and polarization dependence. In section III we perform crystal 
field multiplet calculations to examine the energy dependence of the 
scattering and we discuss the distinct effects of orbital and JT ordering. 
Conclusions are presented in section IV.

\section{Polarization and Azimuthal Angle Dependence.}

In contrast to the $K$ edge experiment, interpretation
of the $L_{II(III)}$ edge experiment, where a $2p$ electron is promoted 
directly into the $3d$ shell, is very clear. At one electron level, if the
$3d_{3z^2-r^2\uparrow}$ orbital is filled, (see figure 2,) the edge itself 
consists of the transition $2p \rightarrow 3d_{x^2-y^2\uparrow}$. This will 
clearly have a very different amplitude if the incoming photon is polarized 
parallel rather than perpendicular to the local ``$z$'' direction 
(${\mathbf\hat z}$) of the ion. This local $z$ direction alternates along the 
$(1,1,0)$ direction between the $a$ and $b$ axes of the crystal, 
with periodicity $2\sqrt{2}$ (relative to the original unit cell). For light 
polarized in the $ab$ plane one therefore anticipates seeing
the $(\frac{\rm 1}{\rm 4},\frac{\rm 1}{\rm 4},0)$ forbidden reflection, the
amplitude being proportional to the difference between the scattering amplitude 
for a Mn$^{3+}$ ion with its local $z$ direction parallel to the crystal $a$ 
axis and the scattering amplitude for one with its local $z$ 
parallel to $b$. This is, 
of course, the same as the difference between the amplitudes for light 
polarized parallel and perpendicular to the $z$ direction of an individual ion. 
Light polarized parallel to the crystal $c$ axis, however, is perpendicular to 
the local $z$ directions of all the Mn$^{3+}$ ions, so the scattering factor 
is the same at each site, and the scattering must be zero. This 
leads to a complex dependence on polarization and on the azimuthal angle around 
the scattering vector. 

More rigorously, the scattering can be viewed as originating in the 3$^{rd}$ 
term of the single ion E1 resonant scattering amplitude given by Hannon et 
al.\cite{Hannon}
\begin{eqnarray}
f_{ion}^{E1} = (\epsv^{f *}_\cdot{\mathbf\hat z})
(\epsv^{0}_\cdot{\mathbf\hat z})
\left(2F_{1,0}^{(e)} - F_{1,1}^{(e)} - F_{1,-1}^{(e)}\right),
\end{eqnarray}
\noindent where $F_{1,q}$ are the spherical components of the transition 
amplitude and $\epsv^{0}$($\epsv^{f}$) the polarization vector
for the incoming (outgoing) beam. The scattering amplitude at the 
$(\frac{\rm 1}{\rm 4},\frac{\rm 1}{\rm 4},0)$ reflection is given by the 
difference between $f_{ion}^{E1}$ for two Mn$^{3+}$ ions with 
${\mathbf\hat z}$ equal to ${\mathbf\hat a}$ and ${\mathbf\hat b}$. 
(Unit vectors along $a$ and $b$ respectively.) Hence
\begin{eqnarray}
f^{E1} = \left[\begin{array}{c}(\epsv^{f *}_\cdot{\mathbf\hat 
a})(\epsv^{0}_\cdot{\mathbf\hat a})\hbox{\hglue 0.4truecm}\\
\hbox{\hglue 0.4truecm} 
- (\epsv^{f *}_\cdot{\mathbf\hat b})(\epsv^{0}_\cdot{\mathbf\hat b})\\ 
\end{array}\right]
\left(2F_{1,0}^{(e)} - F_{1,1}^{(e)} - F_{1,-1}^{(e)}\right).
\end{eqnarray}
The polarization dependence, being purely geometric, is the same as that 
previously observed at the $K$ edge\cite{Murak,Murak2}. Resolving $\epsv$ 
into $\sigma$ and $\pi$ components, and performing two rotations (firstly
through the azimuthal angle $\phi$, secondly through $\frac{\pi}{4}$ around 
${\mathbf\hat c}$, since the wavevector is along $(1,1,0)$ but the orbitals
alternate between $(1,0,0)$ and $(0,1,0)$) we can express the polarization 
in terms of the crystal coordinates. It is then straightforward to show that 
$\sigma^{0} \rightarrow \sigma^{f}$ and $\pi^{0} \rightarrow \pi^{f}$ 
scattering is forbidden. For the $\sigma^{0} \rightarrow \pi^{f}$ and 
$\pi^{0} \rightarrow \sigma^{f}$ channels the scattering intensity turns 
out to be
\begin{eqnarray}
I\left(\theta,\phi\right) = 
\cos^{2}\theta\sin^{2}\phi\left(2F_{1,0}^{(e)} - F_{1,1}^{(e)} - 
F_{1,-1}^{(e)}\right)^{2},
\end{eqnarray}
where $2\theta$ is the scattering angle, and $\phi$ the azimuthal angle around 
the scattering vector.

More interesting is the energy dependence. From the na\"{\i}ve description 
above it is intuitively clear that there must be at least one energy range 
where $I(0,\frac{\pi}{2})\neq 0$, since $3d_{3z^2-r^2\uparrow}$ is filled and 
$3d_{x^2-y^2\uparrow}$ is empty. Indeed, one expects there to be 
scattering in a second, higher energy, as the presence of an 
electron in the 
$3d_{3z^2-r^2\uparrow}$ orbital will split the $3d_{3z^2-r^2\downarrow}$
and $3d_{x^2-y^2\downarrow}$ orbitals by the Coulomb interaction. This 
will happen even in the absence of any JT distortion. This is
because the Coulomb interaction between two electrons 
occupying orbitals with the 
same spatial distribution should be much larger than that between 
electrons in orbitals with different spatial distributions. If the 
latter Coulomb interactions are neglected, then, at one electron level,
we should not expect any splitting in the $t_{2g\downarrow}$ level,
unless it comes from Jahn-Teller effects. We therefore also anticipate 
some differences between the case of orbital ordering alone, and that
of combined orbital and JT ordering. As discussed above, the latter 
case is the most likely, so we would here anticipate three main peaks 
at both the $L_{II}$ and $L_{III}$ edges. 

These arguments tell us nothing
about the relative size or spacing between the peaks, or of the possibility
of smaller peaks being obscured by larger ones.
So, to be more concrete, and to have a more detailed idea of 
the energy dependence that can be expected for $I(0,\frac{\pi}{2})$, 
we have performed an
atomic multiplet calculation for Mn$^{3+}$ in a $D_{4h}$ crystal field, using 
the Cowan multiplet codes and the ``Racah'' crystal field programme of 
B. Searle. There being no clear set of crystal field parameters in the 
literature we first performed a fit to the soft-XAS spectrum for 
LaMnO$_{3}$\cite{Abbate}. The atomic environment of the Mn$^{3+}$ ions in 
LaMnO$_{3}$ is similar in coordination and symmetry to that in our case.

\section{Crystal Field Multiplet Calculations and Energy Dependence.}
\subsection{Fit to the XAS Spectrum of LaMnO$_{3}$.}

Our fit to the soft-XAS spectrum is shown in figure 3a. Hartree-Fock 
values for the Slater integrals are scaled to 65$\%$, and crystal fields 
parameters are $X^{400}$ = 3.42, $X^{420}$ = -4.05 and $X^{220}$ = -2.34, 
where $420$ etc are the relevant branchings for the crystal field group chain 
$O_3 \rightarrow O_h \rightarrow D_{4h}$, in Racah notation. (This corresponds 
to $D_{q} = 0.25$, $D_{s} = 0.28$ $D_{t} = 0.25$ in standard notation.) 
The scaling of the Hartree-Fock parameters is strong, 
but this is in keeping with the findings of previous related studies 
\cite{Pellegrin,Abbate}. Note also that the
line of parameters $D_{s} = 0.55 - D_{q}$, $D_{t} = 2D_{q} - 0.25$, 
$D_{q} = 0.15 \rightarrow 0.25$, with $60\% \rightarrow70\%$ scaling,
produces very similar results.

Using $D_{4h}$ symmetry we find good agreement with the experiment, in contrast 
to Abbate et al.\cite{Abbate} who got only a rough fit using $O_{h}$ symmetry. 
There remain, 
however, a few features of the spectra that do not quite match. These should be due 
partly to the presence of ligand holes, (absent in our calculation) and partly 
to the neglect of the inequivalence between the $x$ and $y$ directions. 
(Each Mn$^{3+}$ has one of these in the crystal's $ab$ plane, the other along 
the $c$ axis.) This would reduce $D_{4h}$ to $D_{2h}$, with an additional 
splitting between $3d_{xz}$ and $3d_{zy}$, and alterations to others.

It is also the case that there is an anisotropy at the Mn$^{4+}$ sites, 
since each Mn$^{4+}$ has two filled Mn$^{3+}$ $3d_{3z^2-r^2}$ 
orbitals pointing towards it, set $90^{\circ}$ apart, and two empty. 
(See figure 1.) This breaks the inversion symmetry and is modulated with the 
same wavevector as the orbital ordering itself. However, the 
Mn$^{3+}$ $3d_{3z^2-r^2}$ lie the other side of the intervening oxygen sites, 
so the effect should be tiny compared to that on the 
Mn$^{3+}$ sites. It should also occur at a slightly different 
energy. We are therefore confident that any effect seen in this experiment 
would be arising from the orderings on the Mn$^{3+}$ sublattice itself.

In figures 3b and c we include the
XAS contributions from X-rays polarized in the $xy$ plane and along the 
$z$-axis. Although the one-electron picture is blurred out by the multiplet 
interactions it is still possible to discern about 4 broad levels, most clearly at the 
$L_{III}$ edge. The 1$^{st}$ and 4$^{th}$ are polarized mostly in the $xy$ 
plane, the 2$^{nd}$ largely parallel to the $z$-axis, but with significant $xy$ 
contributions also.
The 3$^{rd}$ is again mixed, but predominantly $xy$. The first band 
can be reasonably identified as the $3d_{x^2-y^2\uparrow}$ level, albeit rather 
broadened and with other contributions mixed in. The others can probably be 
labelled, at one electron level, according to the scheme in figure 2, provided 
the $3d_{xy\downarrow}$ and $3d_{3z^2-r^2\downarrow}$ are sufficiently
broadened and shifted that they overlap completely.
Hence, the 2$^{nd}$ level comprises mostly the $3d_{xz\downarrow}$
and $3d_{zy\downarrow}$ of the split $t_{2g\downarrow}$ level, and the 3$^{rd}$ 
the $3d_{xy\downarrow}$ component, overlapping with the 
$3d_{3z^2-r^2\downarrow}$ from the $e_{g\downarrow}$.
Finally, the 4$^{th}$ level would be from transitions to 
the $3d_{3x^2-y^2\downarrow}$ orbital. 

\subsection{Resonant X-Ray Scattering.}

Turning to the resonant scattering, we plot in figure 4a the maximum scattering 
intensity, $I(0,\frac{\pi}{2})$. We see that there is a distinct
structure as a function of energy. Comparing the energy scale with
that of figure 3 we note also that the greatest intensity 
does not come from transitions to the empty $3d_{x^2-y^2\uparrow}$ 
orbital itself, but, from transitions to the 
split $t_{2g\downarrow}$ levels, just above. This strong scattering
peak occurs only at the $L_{III}$ edge. On its high energy side
we see a shoulder, but any shoulder to the low energy
side is too small to be noticeable. At the $L_{II}$ edge we see two 
main peaks, with a shoulder on the high energy side of the lower one.
The nature of these other peaks and shoulders will be discussed later.

The energy dependence of $I(0,\frac{\pi}{2})$ carries specific information 
about the environment of the Mn$^{3+}$ sites, helping us answer questions 
to do with the type and origin of the ordering, as discussed 
previously. For example, in the $K$ edge experiments it was not 
possible\cite{Murak} to differentiate between the 
$3d_{3x^2-r^2}$/$3d_{3y^2-r^2}$ orbital ordering 
actually believed to occur and the alternative $3d_{x^2-z^2}$/$3d_{z^2-y^2}$ 
ordering. However, at the $L$ edges the two should have very different energy 
dependences. To illustrate this point, we have recalculated the $L_{II(III)}$ 
edge scattering, keeping the same crystal field magnitudes
as before, but reversing the sign of the $D_{4h}$ terms $X^{420}$ and 
$X^{220}$. This makes $3d_{x^2-y^2}$ the occupied orbital, 
mimicking the alternative $3d_{x^2-z^2}$/$3d_{z^2-y^2}$ ordering. The result 
is shown in figure 4b, and is clearly distinguishable from 4a. 
We emphasize, however, that this curve is not intended as a specific 
prediction, since it is not derived from any experimental spectra for a 
Mn$^{3+}$ ion with the $3d_{x^2-y^2\uparrow}$ orbital filled. It is intended
just as an illustration that much more information should be available at the 
$L$ edges than at the $K$ edge. Extraction of such information would require 
detailed fits to actual experimental data, when such exist. 

\subsection{Distinguishing Orbital Order and Jahn-Teller Order.}

It should also be possible to differentiate between the scattering due to 
orbital ordering alone and that due to combined orbital and JT distortion 
ordering, helping us tackle the question
of which mechanism is the more important. Within the confines of the multiplet
codes we need to keep $X^{420}$ and $X^{220}$ non-zero
in order to make $3d_{3z^2-r^2\uparrow}$ the occupied orbital. This means that 
so far we have actually included the JT effects as well, implicitly assuming 
the involvement of that mechanism. We would now like to identify which parts 
of the predicted 
spectrum, if any, come from the JT ordering, and which come from orbital 
ordering alone. To do this, we note first that whilst $X^{420}$ 
and $X^{220}$ must remain non-zero, in order to split $3d_{3z^2-r^2}$ and 
$3d_{x^2-y^2}$ and observe orbital ordering at all,
the actual size of the splitting required is not important, down to some 
limit set by truncation within the code. Thus we can choose a very 
small tetragonal distortion in order to select the $3d_{3z^2-r^2}$ orbital 
in the initial state, and then use scaling arguments to differentiate between
the orbital ordering effects and the residual JT effects. Hence we can scale 
$X^{420}$ and $X^{220}$ by some $\delta\rightarrow 0$, progressively removing 
the effects of the JT distortions, whilst keeping the scattering 
from the ordered orbitals. (Note that $\delta$ is not intended as an
experimental fitting parameter, it is simply a tool to ``switch off''
the JT distortion, leaving pure orbital ordering, so that we
can separate out the two contributions.)

In figure 5 we show the scattering for a few values 
of $\delta$. (The ratio $\frac{X^{420}}{X^{220}}$ is kept constant.) 
For $\delta < 0.25$ the JT effects are small
and we are essentially left with the effect of the orbital ordering alone.
In figure 6 we plot the heights of the four main peaks against $\delta$.
It is clear that there is only one significant peak due directly to the 
JT distortion. It is labelled with a square symbol on figure 5, and lies
in the $L_{III}$ edge. This is the 
peak corresponding earlier to transitions into the split $t_{2g\downarrow}$ 
levels. In figure 6 the peak height scales to a very small value 
as $\delta\rightarrow0.0$, indicating an OO contribution of only about 4\%.
A better estimate might have come from scaling the
weight under the peak, but this is complicated by the presence of the shoulder.
Comparing the weight under this peak in the $\delta=0.005$ and 
$\delta=1.000$ curves, an OO contribution of around 1.3-1.7\% is obtained, 
depending on where one cuts the shoulder.
That this peak should be due essentially to the JT distortions rather 
than the presence of the ordered $e_{g}$ electron is in complete agreement
with our previous one electron level arguments. 
The equivalent peak at the $L_{II}$ edge is the shoulder on the lower 
main peak. The $\delta$ dependence of its energy is different from that
of the two main peaks at this edge, however, so it is visible as a separate 
peak in the $\delta=0.5$ curve. At smaller $\delta$ it is too small
to be distinguishable. The difference between this peak and the equivalent
peak at the $L_{III}$ edges
is due to the core hole potential; this we have verified by repeating the 
calculation with the core hole potential absent.

The three other peaks are due principally to the ordered orbital 
occupancy, as their heights do not diminish with diminishing $\delta$. Indeed,
the heights are stable over about two orders of magnitude, and scale to 
non-zero values as the JT distortions go to zero. (Their collapse to zero for 
very small $\delta$ is an artifact of numerical truncation in the calculation.) 
 The most interesting of these is the peak labelled with a circle
at the $L_{II}$ edge.
Comparison with the peak identifications in figure 3 (see previous section) 
shows that this is due to resonant transitions into the unoccupied
$3d_{x^2-y^2\uparrow}$, so we would expect to see it even 
in the complete absence of JT distortions. Estimating the size of the
JT contribution from the $\delta$ scaling of the peak height is difficult, but
suggests a negative contribution of around 8-20\%. It is equally difficult to 
use scaling of the weight under the curve, as there is again a shoulder. 
Depending on where we cut the shoulder we get estimates in the 10-30\%
range. It can also be seen clearly that the location of the peak 
shifts downwards in energy by 1.06eV. The equivalent peak at the $L_{III}$
edge moves even further, being invisible for $\delta=1.000$, hidden under 
the square labelled peak. At first glance this movement might 
suggest that the contributions from JT ordering are much larger, but
this is not the case. At one electron level (see figure 2) 
we see that, even in the absence of any JT distortion, this
peak should exist as soon as the $3d_{3z^2-r^2\uparrow}$ orbital is 
occupied and ordered. Any JT distortion on top of this will not
add or take anything at all from the scattering 
intensity, but it will move the $3d_{x^2-y^2\uparrow}$ level upwards in 
energy, and hence also the scattering. This is indeed what we see in figure
5. The changes in peak height and weight come only when multiplet 
contributions beyond one electron level are included. Hence, the location 
of this peak in energy is controlled partly by OO and partly by JT ordering, 
but its existence and weight are still essentially due to the OO itself.

The other two peaks arise from the splitting of the $e_{g\downarrow}$
levels. At the $L_{II}$ edge this peak (labelled with a diamond) 
moves downwards as the JT distortions are switched off, and actually
grows. The equivalent peak at the $L_{III}$ edge is not labelled, as
it is weaker, and moves from being a shoulder on the square labelled
peak at $\delta=1.000$ to being a shoulder on the triangle labelled 
peak at $\delta=0.050$ and below. The diamond labelled peak actually 
grows by almost a factor of
two as the JT distortions are removed. This is again understandable 
at one electron level. As we noted earlier, in the absence of any 
distortion, we would expect the $3d_{3z^2-r^2\downarrow}$ orbital to 
lie above the $3d_{x^2-y^2\downarrow}$ orbital due Coulomb interaction
with the occupied $3d_{3z^2-r^2\uparrow}$. However, in the 
absence of the Coulomb interaction, but with the
JT distortion elongating the ion along the $z$ axis, we would expect
$3d_{3z^2-r^2\downarrow}$ to lie below $3d_{x^2-y^2\downarrow}$.
Hence for the diamond labelled peak the two contributions are in 
competition. Apparently the JT dominates at $\delta=1.000$, since figure
3 indicates that $3d_{3z^2-r^2\downarrow}$ lies below 
$3d_{x^2-y^2\downarrow}$. In figure 6 we see a minimum in the peak 
height around $\delta=0.500$, where the two contributions balance,
(multiplet broadening prevents the peak disappearing completely.)
Below this OO dominates, and the peak intensity grows.

The prediction from the $\delta$ scaling is that the scattering shown in 
figure 4a is dominated at the $L_{III}$ edge by JT ordering, although 
orbital order leads to a clear shoulder to the high 
energy side of the main JT peak. At the $L_{II}$ edge, on the other hand, 
whilst the scattering is predicted to be rather weaker, it is dominated heavily
at the lower end by the orbital ordering. Observation of 
scattering intensity here would thus be a reasonably good measurement of 
orbital order, independent of the presence or absence of JT order. Confirmation of 
this could again be sought by more detailed fitting to experimental data, were 
the measurement to be actually performed. (We anticipate some difference 
between experimental data and that shown in figure 4a since our calculation
in based upon a fit to the XAS spectrum for Mn$^{3+}$ in a slightly 
different setting.)

\subsection{Orbital Ordering in Other Materials.}

Returning now to the other manganite materials, we note that, for example, 
La$_{0.5}$Ca$_{0.5}$MnO$_{3}$ shows exactly the same charge and
orbital ordering in the $ab$ plane\cite{Radaelli}, leading to the same energy, 
polarization and 
azimuthal dependence as for the layered material. The unit cell is normally 
indexed differently, so that the fundamental wavevector is 
$(\frac{\rm 1}{\rm 2},0,0)$, but this again gives a period of about 10.9\AA, 
or an angle of 62.9$^{\circ}$ at the Mn $L_{III}$ edge. 
Similarly for La$_{0.33}$Ca$_{0.67}$MnO$_{3}$, orbital order has been reported 
\cite{Chen} with a wavevector of $(\frac{\rm 1}{\rm 3},0,0)$, giving a 
period around 16.2\AA. The structure is slightly different, but the 
Mn$^{3+}$ local $z$ directions alternate between $(1,1,0)$ and $(1,-1,0)$, 
still giving the same energy, polarization and azimuthal dependence. In 
practise, we expect that this technique, if realised, could measure and 
differentiate between both JT and orbital ordering in a 
wide variety of manganite materials. The exception, unfortunately, is 
LaMnO$_{3}$ itself. Here, in the absence of Mn$^{4+}$ ions, the period of the
orbital order is only about 5.4\AA, too short for the
 Mn $L_{II(III)}$ edges. The minimum orbital order period for which
a reflection could exist at the Mn $L_{III}$ edge is about 9.7\AA.

\section{Conclusion.}

We have demonstrated that, in principle, it should be possible to 
make direct observations of orbital ordering as well as Jahn-Teller ordering
in many of the manganite materials, using resonant X-ray scattering at the 
Mn $L_{II(III)}$ edges. This is likely to be true also of resonant $L$ edge 
scattering in other materials which combine orbital ordering with charge 
ordering. For the current case of the manganites, we have shown that 
sensitivity at the $L_{III}$ edge should be primarily to the accompanying 
Jahn-Teller ordering, whilst that at the $L_{II}$ edge should be due to the 
orbital ordering itself. 
The intensity would have specific energy and polarization dependences, 
and a $\sin^{2}$ dependence on the azimuthal angle around the scattering 
vector. The measurement would be theoretically much more direct than
the disputed resonant X-ray scattering measurements so far performed at the Mn 
$K$ edge, because dipole selection rules allow scattering directly
from the ordered orbitals themselves, rather than from some other unoccupied
orbitals, strongly hybridised with surrounding oxygen orbitals, higher up in 
energy. With the aid of suitable fitting of the energy dependence, the 
measurement
would provide much more detailed information, particularly about the type of 
ordering present, the orbitals actually involved and the relative importance 
of the possible ordering mechanisms.

\begin{acknowledgements}

CWMC would like to thank P. Carra for helpful discussions and 
assistance in learning to use the Cowan and Racah codes. He also thanks E. 
Pellegrin, T. Forgan, 
F. de Bergevin and P. Abbamonte for helpful discussions.
\end{acknowledgements}

\begin{figure}[h]
\vglue -0.4truecm
\centering
\epsfxsize 6.5cm
\leavevmode
\epsfbox{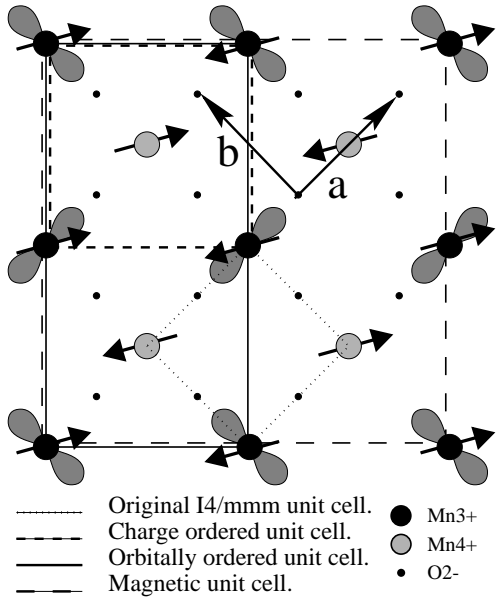}
\vglue 0.1truecm
\caption{Charge, orbital and spin ordering in the MnO$_{2}$ planes of 
La$_{0.5}$Sr$_{1.5}$MnO$_{4}$.}
\end{figure}

\begin{figure}[h]
\vglue -1.7truecm
\centering
\epsfxsize 6.0cm
\leavevmode
\epsfbox{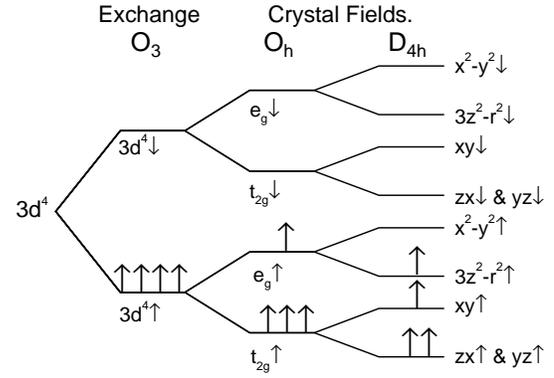}
\vglue -1.8truecm
\caption{Schematic one-electron energy level diagram for the $3d$ shell of 
Mn$^{3+}$ in a tetragonally distorted oxygen octahedron.}
\end{figure}

\begin{figure}[h]
\vglue -0.700truecm
\centering
\epsfxsize 7.1cm
\leavevmode
\epsfbox{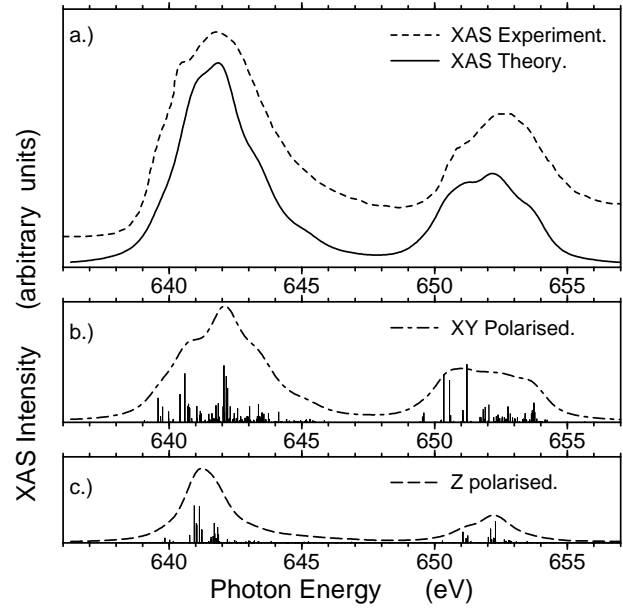}
\vglue -1.55truecm
\caption{Soft-XAS spectrum for Mn$^{3+}$ in LaMnO$_{3}$,
a) Experiment (taken from reference 17) and $D_{4h}$ crystal field multiplet 
calculation. b) $xy$ and c) $z$ polarized contributions.}
\end{figure}

\begin{figure}[h]
\vglue -1.20truecm
\centering
\epsfxsize 7.1cm
\leavevmode
\epsfbox{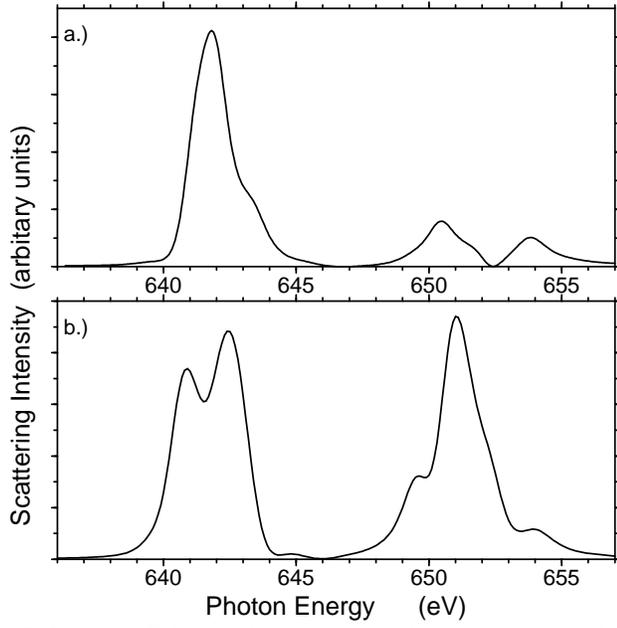}
\vglue -0.82truecm
\caption{a) Calculated intensity at the Mn $L_{II(III)}$ edges. b) Calculated
intensity with signs of $X^{420}$ and  $X^{220}$ reversed.}
\end{figure}

\begin{figure}[h]
\vglue -1.35truecm
\centering
\epsfxsize 7.25cm
\leavevmode
\epsfbox{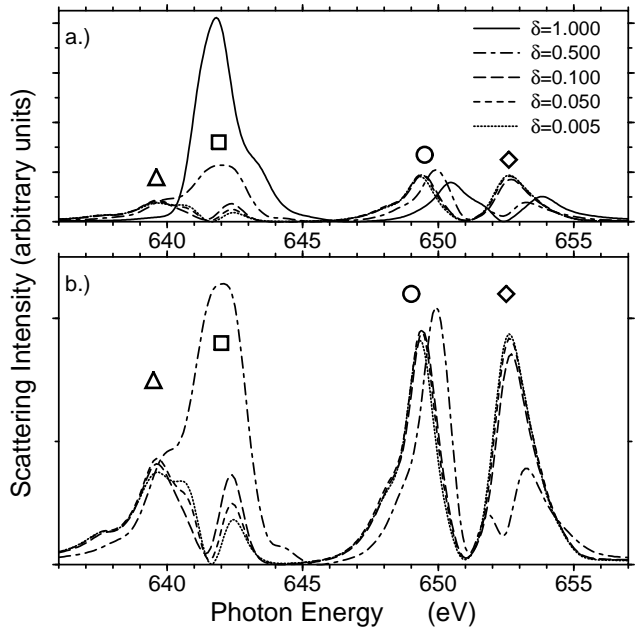}
\vglue -0.70truecm
\caption{
a) Scattering with $D_{4h}$ contributions reduced by $\delta$. Symbols
label the peaks scaled with $\delta$ in figure 6. b) Peak detail at lower
$\delta$ values.}
\end{figure}

\begin{figure}[h]
\vglue -1.35truecm
\centering
\epsfxsize 7.25cm
\leavevmode
\epsfbox{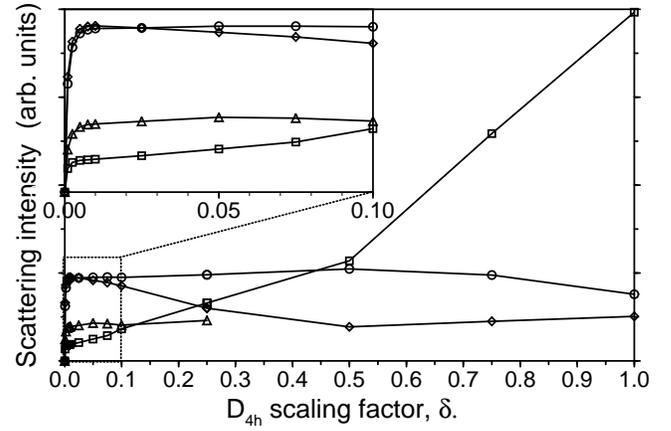}
\vglue -3.00truecm
\caption{
Heights of the peaks labelled by symbols in figure 5, plotted against the 
scaling parameter $\delta$. Inset gives 
$\delta = 0.0\rightarrow 0.1$.}
\end{figure}

\end{document}